# Impact of newly measured $^{26}$Al$(n, p)^{26}$Mg and $^{26}$Al$(n, \alpha)^{23}$Na reaction rates on the nucleosynthesis of $^{26}$Al in stars

Umberto Battino[1,2‹], Claudia Lederer-Woods,[2,3] Marco Pignatari,[4,5,1,2,6] Benjámin Soós,[4,5,7] Maria Lugaro,[4,5,7,8] Diego Vescovi,[9] Sergio Cristallo,[10,11] Philip J Woods[3] and Amanda Karakas[8,12]

[1]*E. A. Milne Centre for Astrophysics, University of Hull, Cottingham Road, Kingston upon Hull, HU6 7RX, UK*
[2]*The NuGrid Collaboration, http://www.nugridstars.org*
[3]*School of Physics and Astronomy, University of Edinburgh, EH9 3FD Edinburgh, UK*
[4]*Konkoly Observatory, Research Centre for Astronomy and Earth Sciences, Eötvös Loránd Research Network (ELKH), Konkoly Thege M. u′t 15-17, H-1121 Budapest, Hungary*
[5]*CSFK, MTA Centre of Excellence, Konkoly Thege Miklo′s u′t 15-17, H-1121 Budapest , Hungary*
[6]*Joint Institute for Nuclear Astrophysics - Center for the Evolution of the Elements, Michigan State University, 640 South Shaw Lane, East Lansing, MI 48824, USA*
[7]*ELTE Eötvös Loránd University, Institute of Physics, Pázmány Péter sétány 1/A, 1117 Budapest, Hungary*
[8]*School of Physics and Astronomy, Monash University, VIC 3800, Australia*
[9]*Goethe University Frankfurt, Max-von-Laue-Strasse 1, Frankfurt am Main 60438, Germany*
[10]*INAF, Observatory of Abruzzo, Via Mentore Maggini snc, 64100 Teramo, Italy*
[11]*Istituto Nazionale di Fisica Nucleare, Sezione di Perugia, 06123 Perugia, Italy*
[12]*Centre of Excellence for Astrophysics in Three Dimensions(ASTRO-3D), Melbourne, Victoria, Australia*

## ABSTRACT

The cosmic production of the short-lived radioactive nuclide $^{26}$Al is crucial for our understanding of the evolution of stars and galaxies. However, simulations of the stellar sites producing $^{26}$Al are still weakened by significant nuclear uncertainties. We re-evaluate the $^{26}$Al$(n, p)^{26}$Mg, and $^{26}$Al$(n, \alpha)^{23}$Na ground state reactivities from 0.01 GK to 10 GK, based on the recent n-TOF measurement combined with theoretical predictions and a previous measurement at higher energies, and test their impact on stellar nucleosynthesis. We computed the nucleosynthesis of low- and high-mass stars using the Monash nucleosynthesis code, the NuGrid mppnp code, and the FUNS stellar evolutionary code. Our low-mass stellar models cover the 2–3 M$_\odot$ mass range with metallicities between $Z = 0.01$ and 0.02, their predicted $^{26}$Al/$^{27}$Al ratios are compared to 62 meteoritic SiC grains. For high-mass stars, we test our reactivities on two 15 M$_\odot$ models with $Z = 0.006$ and 0.02. The new reactivities allow low-mass AGB stars to reproduce the full range of $^{26}$Al/$^{27}$Al ratios measured in SiC grains. The final $^{26}$Al abundance in high-mass stars, at the point of highest production, varies by a factor of 2.4 when adopting the upper, or lower limit of our rates. However, stellar uncertainties still play an important role in both mass regimes. The new reactivities visibly impact both low- and high-mass stars nucleosynthesis and allow a general improvement in the comparison between stardust SiC grains and low-mass star models. Concerning explosive nucleosynthesis, an improvement of the current uncertainties between T9∼0.3 and 2.5 is needed for future studies.

**Key words:** stars: abundances – stars: evolution – nuclear reactions, nucleosynthesis, abundances.

## 1 INTRODUCTION

The short-lived radioactive nuclide $^{26}$Al (with a half-life of 0.72 Myr) is of interest in both γ-ray astrophysics and cosmochemistry, as discussed in details in three recent reviews (Diehl et al. 2021; Diehl 2022; Laird et al. 2022). Its characteristic emission of the diffuse 1809 keV line in our Galaxy detected by γ-ray telescopes (Diehl et al. 1995) is direct evidence for ongoing nucleosynthesis processes enriching the interstellar medium, with a total mass of $^{26}$Al in the Milky Way of nearly 3 M$_\odot$ (Diehl et al. 2006). Moreover, an excess of $^{26}$Mg, the daughter isotope of $^{26}$Al, is observed in meteoritic calcium-aluminium-rich inclusions (CAIs), the first solids to have formed in the protosolar nebula, which provides evidence for injection of live $^{26}$Al in the early Solar System (Lugaro, Ott & Keresztúri 2018). As a consequence, shedding light on the origins of $^{26}$Al is crucial for our understanding of nucleosynthesis processes in stars, the evolution of the Galaxy, as well as the birth of our Solar System.

The stellar production sites of $^{26}$Al in the Galaxy still need to be accurately identified. The spatial distribution of the 1809 keV line suggests that the outflows of Wolf–Rayet stars (M > 25 M$_\odot$ Georgy et al. 2012, Brinkman et al. 2019)) and core-collapse supernovae (CCSNe) are the primary sites of $^{26}$Al production (Prantzos & Casse 1986), accounting for up to about 70 per cent of the live $^{26}$Al detected

‹ E-mail: U.Battino@hull.ac.uk

in the Milky Way (Palacios et al. 2005; Vasini, Matteucci & Spitoni 2022). In particular, $^{26}$Al is produced during three different phases of the evolution of massive stars: (i) H core burning in Wolf–Rayet stars, whose mass-loss is strong enough to eject layers highly enriched in $^{26}$Al located within the H convective core, (ii) explosive C/Ne burning, and (iii) C/Ne convective shell burning during pre-supernova stages, where the fraction of $^{26}$Al that survives the subsequent explosion is then ejected (Limongi & Chieffi 2006; Lawson et al. 2022). Additional $^{26}$Al sources are nova explosions (José, Hernanz & Coc 1997), contributing to up to 30 per cent of the live Galactic $^{26}$Al (Guélin et al. 1995; Bennett et al. 2013) and asymptotic giant branch (AGB) stars, the final phases of low-mass stars (Forestini, Arnould & Paulus 1991), giving an additional small (few per cent) contribution. In all these sites, the direct production mechanism for $^{26}$Al is the well-studied $^{25}$Mg(p, $\gamma$ )$^{26}$Al nuclear reaction (Iliadis et al. 2010; Straniero et al. 2013; Su et al. 2022).

Iliadis et al. (2011) presented a comprehensive investigation of the effects of nuclear reaction rate variations on $^{26}$Al production in massive stars, and listed those nuclear reactions whose uncertainties significantly impact $^{26}$Al synthesis. In particular, they identified the $^{26}$Al(n, p)$^{26}$Mg and $^{26}$Al(n, $\alpha$)$^{23}$Na reactions among the strongest uncertainties impacting the $^{26}$Al abundance. This was due to the scarcity of data for both reactions, with very few previous direct measurements available, and with results highly discrepant. For these reasons, Iliadis et al. recommended $^{26}$Al(n, p)$^{26}$Mg and $^{26}$Al(n, $\alpha$)$^{23}$Na as prime targets for future measurements. This has motivated a new measurement of these reactions at the n_TOF / CERN facility (Lederer-Woods et al. 2021). These $^{26}$Al(n, p)$^{26}$Mg and $^{26}$Al(n, $\alpha$)$^{23}$Na reactions also operate in AGB stars. Indeed, the neutrons provided by the $^{22}$Ne($\alpha$, n)$^{25}$Mg nuclear reaction, activated in the recurring He-flashes (Mowlavi, Jorissen & Arnould 1996; van Raai et al. 2008), trigger $^{26}$Al destruction via $^{26}$Al(n, p)$^{26}$Mg and $^{26}$Al(n, $\alpha$)$^{23}$Na and directly affect the total $^{26}$Al ejected mass.

In this work, we determine new stellar reactivities, including uncertainties, for $^{26}$Al(n, p)$^{26}$Mg and $^{26}$Al(n, $\alpha$)$^{23}$Na. Our evaluation is primarily based on the recent high-precision measurement at the n_TOF-CERN facility, and is supplemented by theoretical calculations and a previous experiment (Trautvetter et al. 1986) at higher neutron energies, to cover the full range of relevant stellar temperatures. The procedure of the evaluation is discussed in Section 2. We apply the new rates in the calculation of full stellar and nucleosynthesis models, and compare the results to key observables in Section 3. Our conclusions are presented in Section 4.

## 2 THE $^{26}$AL(*n*, *p*)$^{26}$MG AND $^{26}$AL(*n*, $\alpha$)$^{23}$NA REACTIVITIES

The $^{26}$Al(n, p)$^{26}$Mg and $^{26}$Al(n, $\alpha$)$^{23}$Na reactivities presented here have been obtained by combining experimental results and theoretical predictions of the respective ground state reaction cross-sections. Up to roughly 150 keV neutron energy, we have used the recent results from n_TOF (Lederer-Woods et al. 2021) for the (n, p) and (n, $\alpha$) cross-sections, respectively. For neutron energies above 150 keV, we determined the cross-sections by theoretical calculations using the Hauser–Feshbach model employed by the nuclear reaction code EMPIRE (Herman et al. 2007). In the case of the (n, p) reaction, we also used previous experimental data obtained at roughly 300 keV neutron energy from the activation measurement of Trautvetter et al. (1986).

The cross-section calculated with EMPIRE depends on assumptions of nuclear level densities, and optical model parameters. The impact of varying model parameters on reaction cross-sections was already studied by Oginni, Iliadis & Champagne (2011). These authors found that different models for nuclear level densities have only a small effect on the reaction cross-sections at the low-neutron energies relevant for this work, while the main variation in the predictions comes from the different choices of optical model parameters. Here, we study the impact of nuclear theoretical uncertainties using the nuclear inputs adopted by Oginni et al. (2011). In particular, we have used Fig. 8 in Oginni et al. (2011) to select the optical model potentials resulting in the minimum and maximum prediction for both, the $^{26}$Al(n, $\alpha$) and the $^{26}$Al(n, p) cross-sections. Specifically, the four sets of optical model potentials used here are defined as: (i) EMPIRE default: Avrigeanu, Hodgson & Avrigeanu (1994) for $\alpha$-particles, and Koning & Delaroche (2003) for protons and neutrons; (ii) Ya-Ko-Hu: Yamamuro (1988) for neutrons, Koning & Delaroche (2003) for protons, and Huizenga & Igo (1962) for $\alpha$-particles; (iii) Ha-Ha-Mc: Harper & Alford (1982) for neutrons and protons and McFadden & Satchler (1966) for $\alpha$-particles; and (iv) Fe-Me-Mc: Ferrer, Carlson & Rapaport (1977) for neutrons, Menet et al. (1971) for protons, and McFadden & Satchler (1966) for $\alpha$-particles.

### 2.1 $^{26}$Al(*n*, *p*)$^{26}$Mg

The cross-section of this reaction has recently been measured with high precision up to neutron energies of 150 keV by Lederer-Woods et al. (2021). These data and associated uncertainties were used to determine upper and lower limits of the $^{26}$Al(n, p)$^{26}$Mg cross-section up to 150 keV neutron energy.

For estimating the corresponding limits at higher neutron energy, we considered previous experimental results and the theoretical EMPIRE calculations. The only available experimental data covering these higher stellar temperatures were obtained by Trautvetter et al. (1986) (with reactivities determined at $T$ = 0.36, 0.82, and 3.6 GK). In the temperature region of overlap ($T$ = 0.36 GK) Trautvetter et al.'s results are lower than those of Lederer-Woods et al., but in agreement within 2 standard deviations. All EMPIRE calculations predict significantly higher reactivities than Trautvetter et al. Hence, to estimate the lower limit of the cross-section from 150 keV to 10 MeV neutron energy, we have scaled the EMPIRE default cross-section to match the experimental reactivity from Trautvetter et al. at T = 3.6 GK. To determine the upper limit of the cross-section above 150 keV, we compare the EMPIRE calculations using different optical model parameters (using the Oginni et al. (2011) inputs) to the experimental cross-sections from Lederer-Woods et al. (2021) between 100 and 150 keV (Fig. 1). This is the energy range where the Hauser–Feshbach approach is predicted to become valid for the $^{26}$Al + n reaction (Rauscher & Thielemann 2000, 2001). The EMPIRE cross-sections are compared to the experimental cross-sections averaged over a large neutron energy range, integrating over several resonances, rather than comparing individual data points which still show resonant structures. The cross-section average of the experimental data from 100 to 150 keV, including systematic and statistical uncertainty is 147 ± 27 mb. Taking an upper limit, for example, of average value plus 1$\sigma$, we obtain a value of 174 mb. This is smaller than the lowest EMPIRE prediction Ha-Ha-Mc which results in ≈240 mb. Here, we adopt the Ha-Ha-Mc calculation as an upper limit for the cross-section from 150 keV to 10 MeV.

Using the procedure described earlier, we obtain lower and upper limits of the cross-section over the entire neutron energy range of interest, which were then used to calculate lower and upper limits of the astrophysical reactivities. The top panel of Fig. 2 shows the resulting reactivity used in our stellar models, with the upper and lower limits, from 0.01 to 10 GK. The reactivity is entirely determined by results from Lederer-Woods et al. (2021) up to temperatures of around 0.4 GK, while from about 2 to 3 GK the upper limit corresponds to the predicted reactivity derived using the Ha-Ha-Mc theoretical calculation. The bottom panel of the Figure shows a comparison to previous experimental data by Koehler et al. (1997), and theoretical predictions by Caughlan & Fowler (1988) used for comparison in the AGB model calculations of Section 3. The recommended reactivities, including upper and lower limits from 0.01 to 10 GK stellar temperature are listed in Table 1.

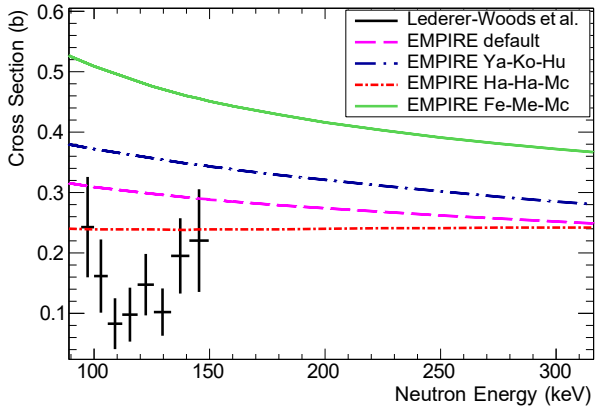

**Figure 1.** The $^{26}$Al$(n, p)^{26}$Mg cross-section from 100 to 300 keV laboratory neutron energy. The results from Lederer-Woods et al. with statistical uncertainties (Lederer-Woods et al. 2021) are compared to predictions of the cross-section using the EMPIRE code with our different combinations of optical model potentials for neutron, proton, and α-particles (see text for details).

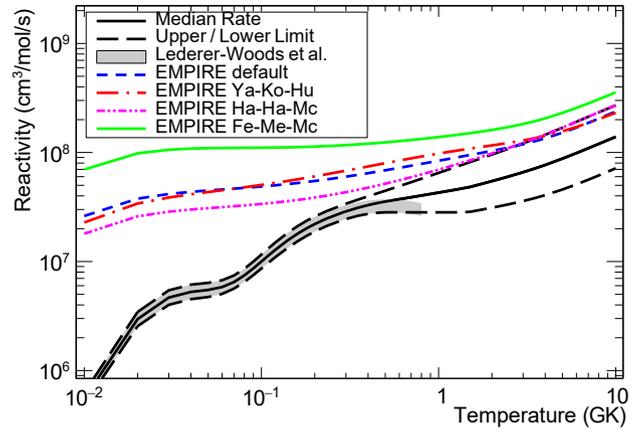

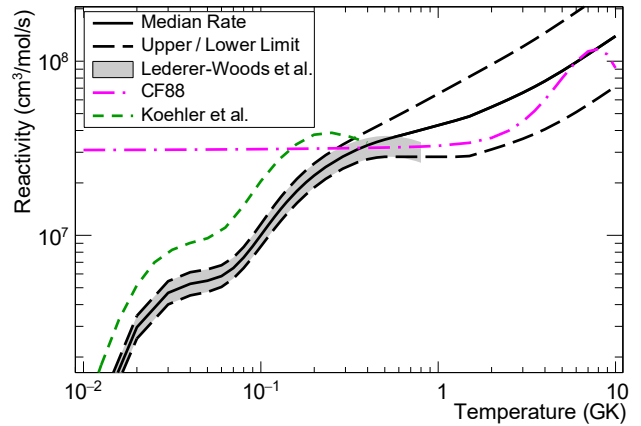

**Figure 2.** (Top) Our $^{26}$Al$(n, p)^{26}$Mg reactivity, as compared to experimental results by Lederer-Woods et al. (2021), and the theoretical cross-sections obtained with the nuclear reaction code EMPIRE for our four different combinations of neutron-, proton- and α- optical model potentials. (Bottom) Our $^{26}$Al$(n, p)^{26}$Mg reactivity compared to Lederer-Woods et al., Koehler et al. (1997), and Caughlan & Fowler (1988).

### 2.2 $^{26}$Al$(n, α)^{23}$Na

Similar to the $(n, p)$ reaction, we have used data from Lederer-Woods et al. (2021) and associated uncertainties to determine upper and lower limits of the $^{26}$Al$(n, p)^{26}$Mg cross-section, up to 160 keV neutron energy. For cross-sections above this energy, we used the EMPIRE predictions, based on the Oginni et al. (2011) inputs.

The determination of the upper and lower limits of the cross-section above 160 keV is described further. Fig. 3 shows the experimental cross-section by Lederer-Woods et al. (2021) from 100 to 160 keV compared to the cross-sections calculated with the four different combinations of optical model potentials. All the models underestimate the cross-section in this energy region. To estimate the lower limits of the cross-section from 160 keV to 10 MeV, we used the lowest prediction, that is, Ha-Ha-Mc. To estimate an upper limit of the cross-section above 160 keV neutron energy, we again compare averaged cross-section values. The experimental cross-section average from 100 to 160 keV including statistical and systematic uncertainties is 115 ± 19 mb, taking the average values plus 1σ as upper limit we obtain 134 mb. The corresponding value for the highest EMPIRE prediction Ya-Ko-Hu is 90 mb, a factor 1.5 smaller. For our upper limit of the cross-section above 160 keV we adopt the Ya-Ko-Hu cross-section scaled by a factor 1.5.

The upper and lower limits of the cross-section obtained as described earlier were then used to calculate upper and lower limits of the stellar reactivities from 0.01 to 10 GK stellar temperature (Table 2). Fig. 4 shows our new rate compared to the theoretical predictions, and experimental results by Lederer-Woods et al. in the top panel of Fig. 4. The bottom panel displays the same as above, but compared to previous experimental results by De Smet et al. (2007)

and recommended cross-sections from the NACRE (Angulo et al. 1999) compilation, both of which have been used for comparison in the AGB model calculations presented in Section 3.

## 3 IMPACT ON NUCLEOSYNTHESIS CALCULATIONS

We tested the impact of our new reactivities by simulating the nucleosynthesis in AGB stars with initial mass M = 2 and 3 M$_\odot$ and metallicities Z = 0.01, 0.014, 0.0167, and 0.02, and in massive stars of M = 15 M$_\odot$ and Z = 0.006 and 0.02.

### 3.1 Low-mass AGB star

In AGB stars, neutron-induced reactions are affected by the uncertainties on the production of neutrons: the uncertainty of the $^{22}$Ne$(α, n)^{25}$Mg reaction and the uncertainties of the stellar physics that control the temperature. The latter uncertainties are mostly related to the modelling of mixing processes in the deep interior of the star. As these are implemented in different ways by different stellar evolution codes, we considered their impact qualitatively by computing the AGB nucleosynthesis using the three different sets of stellar models

**Table 1.** $^{26}$Al(*n*, *p*)$^{26}$Mg reactivities on the ground state of $^{26}$Al in units of [cm$^3$/mol s].

| T [GK] | Lower limit | Median rate | Upper limit |
|---|---|---|---|
| 0.01 | 4.89E + 05 | 5.54E + 05 | 6.28E + 05 |
| 0.02 | 2.55E + 06 | 2.97E + 06 | 3.45E + 06 |
| 0.03 | 4.01E + 06 | 4.67E + 06 | 5.45E + 06 |
| 0.04 | 4.52E + 06 | 5.27E + 06 | 6.14E + 06 |
| 0.05 | 4.72E + 06 | 5.49E + 06 | 6.37E + 06 |
| 0.06 | 5.05E + 06 | 5.83E + 06 | 6.73E + 06 |
| 0.07 | 5.65E + 06 | 6.49E + 06 | 7.45E + 06 |
| 0.08 | 6.49E + 06 | 7.46E + 06 | 8.57E + 06 |
| 0.09 | 7.51E + 06 | 8.66E + 06 | 9.99E + 06 |
| 0.10 | 8.63E + 06 | 1.00E + 07 | 1.16E + 07 |
| 0.11 | 9.81E + 06 | 1.14E + 07 | 1.33E + 07 |
| 0.12 | 1.10E + 07 | 1.28E + 07 | 1.50E + 07 |
| 0.13 | 1.21E + 07 | 1.42E + 07 | 1.67E + 07 |
| 0.14 | 1.33E + 07 | 1.56E + 07 | 1.82E + 07 |
| 0.15 | 1.43E + 07 | 1.68E + 07 | 1.97E + 07 |
| 0.16 | 1.53E + 07 | 1.80E + 07 | 2.11E + 07 |
| 0.17 | 1.63E + 07 | 1.91E + 07 | 2.24E + 07 |
| 0.18 | 1.72E + 07 | 2.02E + 07 | 2.37E + 07 |
| 0.19 | 1.80E + 07 | 2.12E + 07 | 2.48E + 07 |
| 0.20 | 1.88E + 07 | 2.21E + 07 | 2.59E + 07 |
| 0.25 | 2.22E + 07 | 2.59E + 07 | 3.03E + 07 |
| 0.30 | 2.46E + 07 | 2.89E + 07 | 3.39E + 07 |
| 0.35 | 2.62E + 07 | 3.11E + 07 | 3.69E + 07 |
| 0.40 | 2.73E + 07 | 3.29E + 07 | 3.96E + 07 |
| 0.45 | 2.80E + 07 | 3.43E + 07 | 4.22E + 07 |
| 0.50 | 2.82E + 07 | 3.55E + 07 | 4.46E + 07 |
| 0.55 | 2.83E + 07 | 3.64E + 07 | 4.69E + 07 |
| 0.60 | 2.83E + 07 | 3.72E + 07 | 4.91E + 07 |
| 0.65 | 2.83E + 07 | 3.81E + 07 | 5.13E + 07 |
| 0.70 | 2.83E + 07 | 3.88E + 07 | 5.34E + 07 |
| 0.75 | 2.83E + 07 | 3.96E + 07 | 5.54E + 07 |
| 0.80 | 2.83E + 07 | 4.03E + 07 | 5.75E + 07 |
| 0.85 | 2.83E + 07 | 4.10E + 07 | 5.94E + 07 |
| 0.90 | 2.83E + 07 | 4.16E + 07 | 6.13E + 07 |
| 0.95 | 2.83E + 07 | 4.23E + 07 | 6.32E + 07 |
| 1.00 | 2.83E + 07 | 4.29E + 07 | 6.50E + 07 |
| 1.25 | 2.83E + 07 | 4.56E + 07 | 7.36E + 07 |
| 1.50 | 2.85E + 07 | 4.82E + 07 | 8.15E + 07 |
| 1.75 | 2.99E + 07 | 5.15E + 07 | 8.88E + 07 |
| 2.00 | 3.12E + 07 | 5.46E + 07 | 9.57E + 07 |
| 2.25 | 3.24E + 07 | 5.75E + 07 | 1.02E + 08 |
| 2.50 | 3.35E + 07 | 6.04E + 07 | 1.09E + 08 |
| 2.74 | 3.46E + 07 | 6.31E + 07 | 1.15E + 08 |
| 3.00 | 3.59E + 07 | 6.59E + 07 | 1.21E + 08 |
| 3.25 | 3.70E + 07 | 6.87E + 07 | 1.27E + 08 |
| 3.50 | 3.82E + 07 | 7.14E + 07 | 1.33E + 08 |
| 3.75 | 3.94E + 07 | 7.41E + 07 | 1.39E + 08 |
| 4.00 | 4.07E + 07 | 7.69E + 07 | 1.45E + 08 |
| 5.00 | 4.57E + 07 | 8.79E + 07 | 1.69E + 08 |
| 6.00 | 5.10E + 07 | 9.88E + 07 | 1.91E + 08 |
| 7.00 | 5.63E + 07 | 1.09E + 08 | 2.13E + 08 |
| 8.00 | 6.15E + 07 | 1.20E + 08 | 2.33E + 08 |
| 9.00 | 6.66E + 07 | 1.30E + 08 | 2.52E + 08 |
| 10.0 | 7.16E + 07 | 1.39E + 08 | 2.71E + 08 |

by (1) Battino et al. (2019), computed by the NUGRID collaboration with the MESA code (Paxton et al. 2010); (2) Vescovi et al. (2021), computed with the FUNS code (Straniero, Gallino & Cristallo 2006); and (3) Karakas & Lugaro (2016), computed with the MONASH code (Frost & Lattanzio 1996). In all cases, we compared the results to those obtained adopting the $^{26}$Al(*n*, *p*)$^{26}$Mg and $^{26}$Al(*n*, α)$^{23}$Na reactivities from the REACLIB data base (Cyburt et al. 2010), from

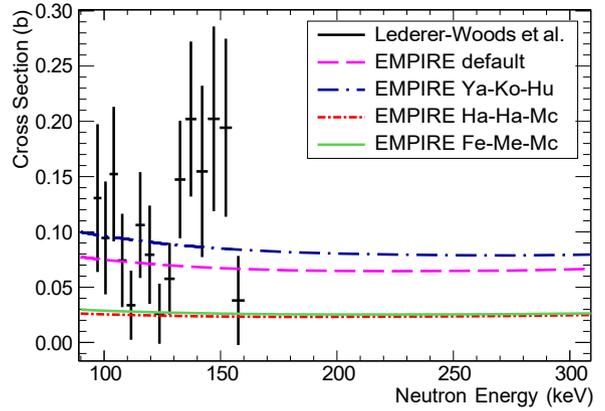

**Figure 3.** The $^{26}$Al(*n*, α)$^{26}$Mg cross-section from 100 to 300 keV laboratory neutron energy. Results from Lederer-Woods et al. (2021) with statistical uncertainties are compared to predictions of the cross-section using the EMPIRE code with our four different combinations of optical model potentials for neutron, proton, and α-particles.

Caughlan & Fowler (1988) and the NACRE compilation (Angulo et al. 1999), respectively. In relation to the additional reactions that have an effect on the nucleosynthesis of $^{26}$Al in AGB stars, we adopted the same rates in all models as follows: $^{25}$Mg(*p*, γ)$^{26}$Al from Straniero et al. (2013), $^{26}$Al(*p*, γ)$^{27}$Si from Iliadis et al. (2010), and $^{22}$Ne(α, *n*)$^{26}$Mg from Adsley et al. (2021). As mentioned earlier, the last reaction has a strong effect on the neutron production and the rate adopted here is lower than those previously used in some of our models (e.g. the rates of Iliadis et al. (2010) were used in the Monash models earlier). This change resulted in an increase of roughly a factor of two in the predicted $^{26}$Al/$^{27}$Al ratios.

The $^{26}$Al/$^{27}$Al measured ratio in mainstream silicon carbide (SiC) grains (Groopman et al. 2015; Liu et al. 2021) represents strong evidence for the production of $^{26}$Al in low-mass AGB stars (1.5 ≤ M$_\odot$ ≤ 4) with around solar metallicity, since these grains are known to be formed in the C-rich winds of these stars (Lugaro et al. 1999, 2003). Fig. 5 shows the FUNS and Monash prediction for the Al and C isotopic ratios, as compared to the SiC data. The theoretical results obtained adopting our recommended (i.e. median), upper, and lower limits of both the $^{26}$Al(*n*, *p*)$^{26}$Mg and $^{26}$Al(*n*, α)$^{23}$Na reactivities are shown in the figure. The impact of our new rates is visible in all models, shifting the theoretical tracks towards higher $^{26}$Al/$^{27}$Al values, as compared to models run adopting the REACLIB reactivities. The same increasing trend occurs relative to models run adopting the $^{26}$Al(*n*, *p*)$^{26}$Mg and $^{26}$Al(*n*, α)$^{23}$Na reactivities from Koehler et al. (1997) and Wagemans et al. (2001), respectively, whose results are close to those obtained when adopting our lower limits values. For both the reactions considered here, the impact of our new rates on AGB nucleosynthesis is larger than their uncertainties: in fact, the theoretical tracks calculated using the values from both REACLIB and Koehler et al. and Wagemans et al., lie outside the $^{26}$Al/$^{27}$Al range covered by those calculated using our upper and lower limits.

The same comparison and similar results is shown in Fig. 6, except for NuGRID and FUNS models at slightly super-solar metallicity (Z = 0.02). In this case, however, a clear difference is visible in the range of $^{26}$Al/$^{27}$Al covered by the two stellar codes. As discussed in Battino et al. (2019), NuGRID models include mixing at the convective boundary at the bottom of the He-intershell during each thermal-pulse event. This favours the mixing of carbon from the stellar core into the intershell and results in a higher and lower abundance of carbon and helium, respectively, as compared to both FUNS and Monash models, which do not include such mixing. Due to the lower helium abundance, NuGRID models require higher temperatures to trigger He-flash episodes, which leads to a stronger activation of the $^{22}$Ne$(\alpha, n)^{26}$Mg nuclear reaction and hence to a higher neutron density. This makes the destruction of $^{26}$Al via both $(n, p)$ and $(n, \alpha)$ channels more effective, decreasing the $^{26}$Al/$^{27}$Al ratio.

**Table 2.** $^{26}$Al$(n, \alpha)^{23}$Na reactivities on the ground state of $^{26}$Al in units of [cm$^3$/mol s].

| T [GK] | Lower limit | Median rate | Upper limit |
|---|---|---|---|
| 0.01 | 9.02E + 05 | 9.92E + 05 | 1.09E + 06 |
| 0.02 | 8.25E + 06 | 9.08E + 06 | 1.00E + 07 |
| 0.03 | 1.34E + 07 | 1.48E + 07 | 1.63E + 07 |
| 0.04 | 1.52E + 07 | 1.67E + 07 | 1.84E + 07 |
| 0.05 | 1.53E + 07 | 1.68E + 07 | 1.85E + 07 |
| 0.06 | 1.47E + 07 | 1.62E + 07 | 1.78E + 07 |
| 0.07 | 1.40E + 07 | 1.54E + 07 | 1.70E + 07 |
| 0.08 | 1.34E + 07 | 1.47E + 07 | 1.62E + 07 |
| 0.09 | 1.29E + 07 | 1.41E + 07 | 1.55E + 07 |
| 0.10 | 1.24E + 07 | 1.36E + 07 | 1.50E + 07 |
| 0.11 | 1.21E + 07 | 1.33E + 07 | 1.46E + 07 |
| 0.12 | 1.18E + 07 | 1.30E + 07 | 1.43E + 07 |
| 0.13 | 1.16E + 07 | 1.27E + 07 | 1.40E + 07 |
| 0.14 | 1.14E + 07 | 1.25E + 07 | 1.38E + 07 |
| 0.15 | 1.12E + 07 | 1.24E + 07 | 1.37E + 07 |
| 0.16 | 1.11E + 07 | 1.23E + 07 | 1.35E + 07 |
| 0.17 | 1.10E + 07 | 1.22E + 07 | 1.34E + 07 |
| 0.18 | 1.09E + 07 | 1.21E + 07 | 1.34E + 07 |
| 0.19 | 1.08E + 07 | 1.20E + 07 | 1.33E + 07 |
| 0.20 | 1.07E + 07 | 1.20E + 07 | 1.34E + 07 |
| 0.25 | 1.06E + 07 | 1.19E + 07 | 1.35E + 07 |
| 0.30 | 1.06E + 07 | 1.22E + 07 | 1.39E + 07 |
| 0.35 | 1.09E + 07 | 1.26E + 07 | 1.47E + 07 |
| 0.40 | 1.11E + 07 | 1.32E + 07 | 1.57E + 07 |
| 0.45 | 1.14E + 07 | 1.39E + 07 | 1.69E + 07 |
| 0.50 | 1.17E + 07 | 1.45E + 07 | 1.80E + 07 |
| 0.55 | 1.20E + 07 | 1.52E + 07 | 1.94E + 07 |
| 0.60 | 1.22E + 07 | 1.59E + 07 | 2.06E + 07 |
| 0.65 | 1.24E + 07 | 1.65E + 07 | 2.20E + 07 |
| 0.70 | 1.26E + 07 | 1.71E + 07 | 2.32E + 07 |
| 0.75 | 1.27E + 07 | 1.76E + 07 | 2.44E + 07 |
| 0.80 | 1.29E + 07 | 1.82E + 07 | 2.56E + 07 |
| 0.85 | 1.30E + 07 | 1.87E + 07 | 2.68E + 07 |
| 0.90 | 1.31E + 07 | 1.91E + 07 | 2.79E + 07 |
| 0.95 | 1.32E + 07 | 1.96E + 07 | 2.91E + 07 |
| 1.00 | 1.33E + 07 | 2.00E + 07 | 3.01E + 07 |
| 1.25 | 1.37E + 07 | 2.20E + 07 | 3.53E + 07 |
| 1.50 | 1.41E + 07 | 2.38E + 07 | 4.00E + 07 |
| 1.75 | 1.47E + 07 | 2.56E + 07 | 4.46E + 07 |
| 2.00 | 1.53E + 07 | 2.74E + 07 | 4.91E + 07 |
| 2.25 | 1.61E + 07 | 2.94E + 07 | 5.36E + 07 |
| 2.50 | 1.70E + 07 | 3.14E + 07 | 5.82E + 07 |
| 2.74 | 1.80E + 07 | 3.36E + 07 | 6.27E + 07 |
| 3.00 | 1.91E + 07 | 3.60E + 07 | 6.77E + 07 |
| 3.25 | 2.04E + 07 | 3.85E + 07 | 7.27E + 07 |
| 3.50 | 2.17E + 07 | 4.11E + 07 | 7.78E + 07 |
| 3.75 | 2.32E + 07 | 4.39E + 07 | 8.30E + 07 |
| 4.00 | 2.48E + 07 | 4.68E + 07 | 8.83E + 07 |
| 5.00 | 3.18E + 07 | 5.95E + 07 | 1.11E + 08 |
| 6.00 | 4.01E + 07 | 7.37E + 07 | 1.36E + 08 |
| 7.00 | 4.93E + 07 | 8.92E + 07 | 1.61E + 08 |
| 8.00 | 5.93E + 07 | 1.05E + 08 | 1.88E + 08 |
| 9.00 | 6.97E + 07 | 1.22E + 08 | 2.14E + 08 |
| 10.0 | 8.05E + 07 | 1.39E + 08 | 2.41E + 08 |

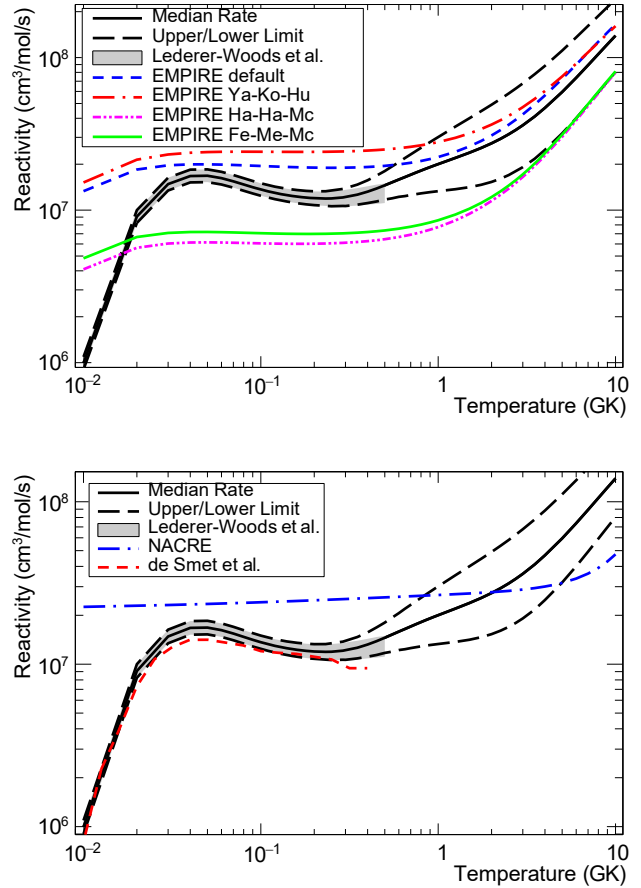

**Figure 4.** (Top) Our $^{26}$Al$(n, \alpha)^{23}$Na reactivity, as compared to experimental results by Lederer-Woods et al., and the theoretical cross-sections obtained with the nuclear reaction code EMPIRE for our four different combinations of neutron-, proton- and $\alpha$- optical model potentials. (Bottom) Our $^{26}$Al$(n, \alpha)^{26}$Mg reactivity compared to Lederer-Woods et al., De Smet et al. (2007), and the rate recommended in the NACRE compilation (Angulo et al. 1999).

Finally, we note that all the stellar models here discussed predict an increase of $^{12}$C/$^{13}$C ratio in the envelope during TDUs higher for 3 than the 2 M$_\odot$ models, consistent with previous literature results (see, e.g. Wasserburg, Boothroyd & Sackmann (1995), Zinner et al. (2006)). In general, such an increase is higher than the $^{12}$C/$^{13}$C ratio measured in SiC grains. It might be an indication of the inclusion of extra-mixing processes, such as the cool bottom process (CBP, see Nollett, Busso & Wasserburg (2003), Zinner et al. (2006), Palmerini et al. (2011)), in which case also $^{26}$Al may be mildly affected, depending on the exact CBP parameters (Nollett et al. 2003). It may also indicate that the parent stars were born with different initial $^{12}$C/$^{13}$C ratio, as expected from the effect of galactic chemical evolution for different metallicities.

In Fig. 7, we show the whole range of $^{26}$Al/$^{27}$Al covered by the Monash, FUNS, and MESA models when adopting our new rates, as compared to the range measured in SiC grains. The theoretical $^{26}$Al/$^{27}$Al values shown represent the combined contribution of 2 and 3 M$_\odot$ at the same metallicity. It's important to notice how the isotopic ratio data range by Liu et al. (2021) is fully covered by the measurements from Groopman et al. (2015). In particular, the lowest $^{26}$Al/$^{27}$Al measured in the two data sets is almost identical. Overall, comparing Monash models at $Z = 0.01$ and $Z = 0.14$, and FUNS models at $Z = 0.0167$ and $Z = 0.02$, the $^{26}$Al/$^{27}$Al ratio increases with metallicity. This is due to the fact that higher metallicity models have a higher initial abundance of $^{26}$Mg, which acts as seed for the production of $^{26}$Al, and that these models are colder, therefore, the neutron source reaction $^{22}$Ne($\alpha, n$)$^{26}$Mg is less activated. As discussed earlier, the NuGrid results are visibly different from those obtained with FUNS models at the same metallicity due to the different treatment of mixing. While Fig. 7 shows that the models can broadly match the measured range, the comparison of the models with the SiC data is currently hampered by systematic uncertainties, not reported in the plots, of the order of a factor of two. This is due to the fact that two different elements need to be measured, Al and Mg, because $^{26}$Mg, the daughter of $^{26}$Al, is needed to derive the initial, now extinct, $^{26}$Al abundance in each grain. A sensitivity factor is therefore introduced in the derivation of the $^{26}$Al/$^{27}$Al ratio due to the different response of the instrument to different elements (Groopman et al. 2015; Liu et al. 2021). These systematic uncertainties from the data add up to those from the stellar models (mostly the temperature controlled by the mixing, as discussed earlier) and the rate of the $^{22}$Ne($\alpha, n$)$^{26}$Mg reaction, which directly affects the $^{26}$Al depletion (Adsley et al. 2021; Ota et al. 2021).

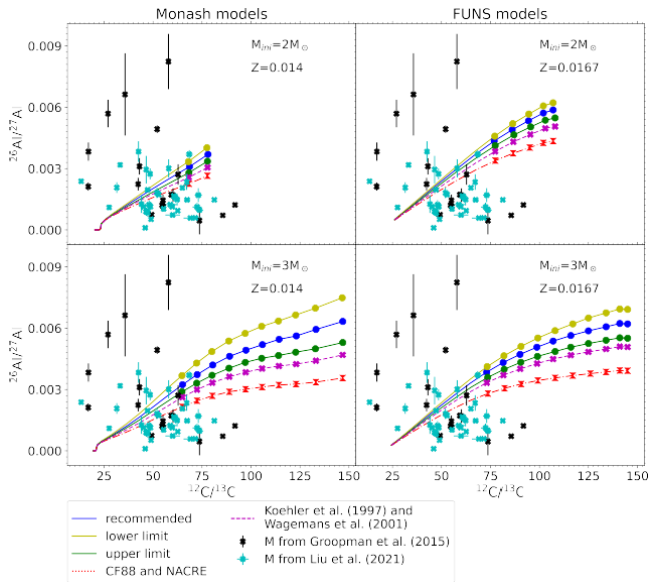

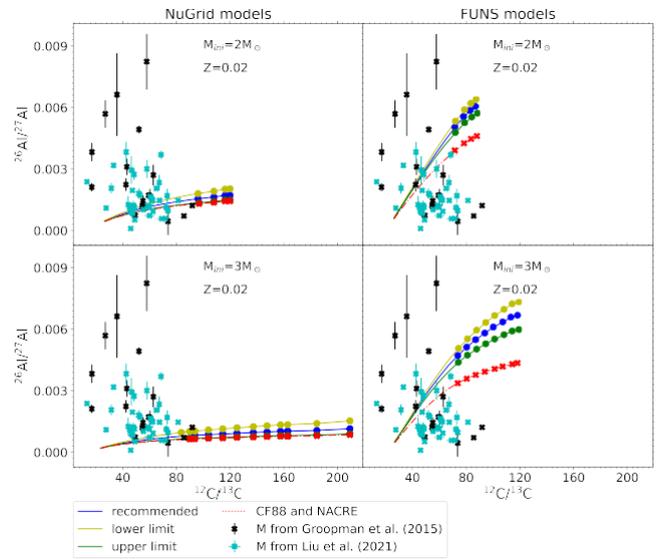

**Figure 5.** Comparison of measured $^{26}$Al/$^{27}$Al and $^{12}$C/$^{13}$C ratios from presolar SiC grains from Groopman et al. (2015) and Liu et al. (2021) with the theoretical predictions of Monash and FUNS low-mass AGB models at solar metallicity. Note that the solar metallicity is $Z = 0.014$ in the Monash models, following Asplund et al. (2009), and $Z = 0.0167$ in the FUNS models, following Lodders (2021) after including the effect of diffusion (Vescovi et al. 2020). Each symbol on the stellar evolution lines marks a TDU event during the C-rich AGB phase, that is, they represent the composition at the time when the conditions for SiC grains condensation are met). The theoretical results obtained adopting our recommended (median) values and our upper and lower limits of both the $^{26}$Al($n, p$)$^{26}$Mg and $^{26}$Al($n, \alpha$)$^{23}$Na reactivities are shown, as well as the predictions obtained with the reactivities recommended by Caughlan & Fowler (1988) and the NACRE compilation (Angulo et al. 1999), and those by Koehler et al. (1997) and Wagemans et al. (2001).

**Figure 6.** Same as Fig. 5, except showing NuGrid and FUNS AGB models at metallicity $Z = 0.02$.

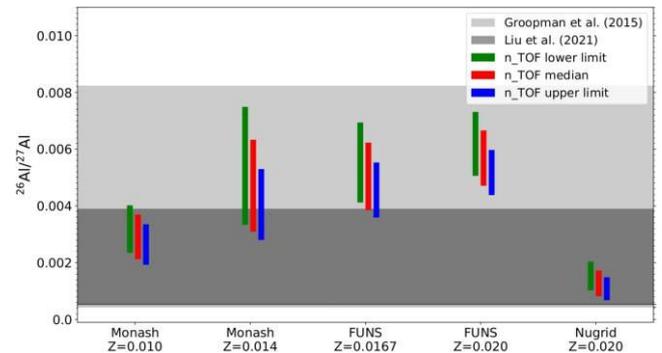

**Figure 7.** Full range of $^{26}$Al/$^{27}$Al ratios predicted by our 2 and 3 M$_\odot$ Monash, FUNS, and MESA models adopting our new $^{26}$Al($n, p$)$^{26}$Mg and $^{26}$Al($n, \alpha$)$^{23}$Na reactivities. Gray bands represent observed $^{26}$Al/$^{27}$Al ratios in presolar grains by Groopman et al. (2015) and Liu et al. (2021). Consistently with Figs 5 and 6, notice how the isotopic ratio data range by Liu et al. (2021) is fully covered by the measurements from Groopman et al. (2015).

### 3.2 Massive stars

Massive stars are the dominant source of $^{26}$Al in the Galaxy (e.g. Timmes et al. 1995; Diehl et al. 2021; Vasini et al. 2022) through stellar winds of Wolf–Rayet stars (e.g. Prantzos & Casse 1986; Meynet et al. 1997; Voss et al. 2009) and CCSNe ejecta (e.g. Timmes et al. 1995; Limongi & Chieffi 2006; Lawson et al. 2022). The bulk of the $^{26}$Al present in massive star winds is made in H-burning conditions, where no relevant neutron source reactions are activated. In contrast, ($n, p$) and ($n, \alpha$) reactions on $^{26}$Al can affect yields from CCSNe. In order to clarify the impact of these two rates on stellar CCSN predictions, we discuss here two different models by Ritter et al. (2018), with initial mass M = 15 M$_\odot$ and two metallicities, $Z = 0.02$ and $Z = 0.006$, respectively. Fig. 8 shows the abundance profiles of the CCSN ejecta for these two cases ($Z = 0.02$ upper panel, $Z = 0.006$ lower panel). As expected, the $^{26}$Al abundance changes by orders of magnitude in the different parts of the CCSN ejecta. We first discuss the hottest parts of the ejecta, and then we move outwards to external layers. Briefly, during O-burning and Si-burning conditions $^{26}$Al is not made and eventually any ashes from previous stages are quickly depleted (at mass coordinates M < 1.8 M$_\odot$ and M < 2.4 M$_\odot$, in the upper and lower panels, respectively). During hydrostatic convective C-burning and explosive C/Ne burning, $^{26}$Al is efficiently produced by proton capture on the abundant $^{25}$Mg. Protons are directly made by C-fusion reactions, and $^{25}$Mg is mostly made by the α-capture on $^{22}$Ne and the neutron capture on $^{24}$Mg. On the other hand, in a C-burning environment the neutron-capture reactions depleting $^{26}$Al can be activated, where the $^{22}$Ne(α, n)$^{25}$Mg reaction is the dominant neutron source. The final $^{26}$Al yields in C-burning ejecta will be given by the interplay between the production and destruction nucleosynthesis channels mentioned earlier. Comparing now the two models with different metallicity, we find that for the case of the Z = 0.02 model (upper panel of Fig. 8) the C-burning ejecta in the mass region 1.8 M$_\odot$ < M < 3 M$_\odot$ are dominated by the pre-explosive C shell production, with a marginal explosive production at M = 1.9 M$_\odot$. For the Z = 0.006 model the explosive production peak at M = 2.65 M$_\odot$ is completely dominating the $^{26}$Al yields from the former C-shell material (M < 3.2 M$_\odot$) and all the $^{26}$Al ejecta of the model.

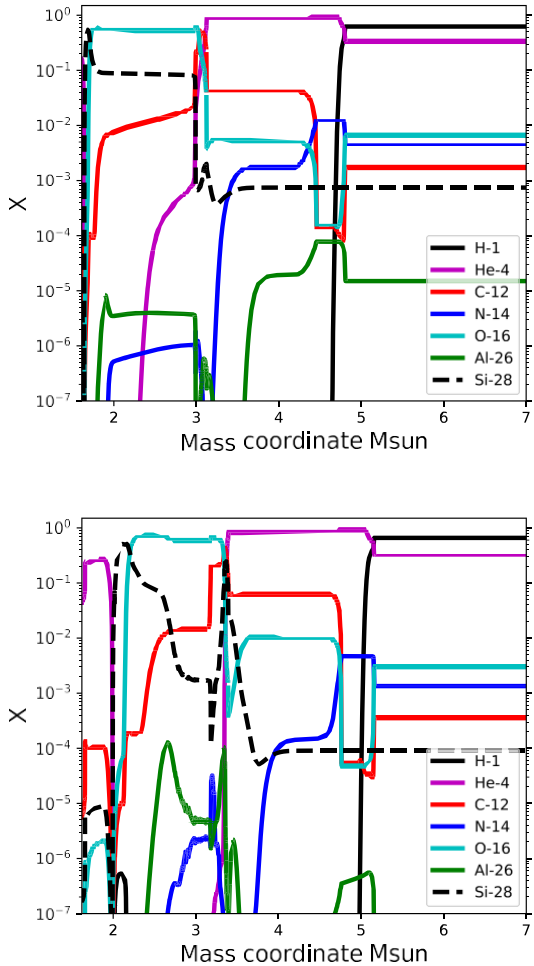

**Figure 8.** Isotopic abundances (mass fractions) with respect to mass coordinate of H, $^4$He, $^{12}$C, $^{14}$N, $^{16}$O, $^{28}$Si, and $^{26}$Al are shown for the models M = 15 M$_\odot$ models at Z = 0.02 (upper panel) and Z = 0.006 (lower panel) after the CCSN explosion (Ritter et al. 2018).

Now we consider the contribution of $^{26}$Al abundances in the external stellar layers to the ejecta. The Z = 0.02 model shows $^{26}$Al non-negligible abundances from mass coordinates ≈3.5 M$_\odot$ outwards. The $^{26}$Al abundance up to 4.4 M$_\odot$ was present in the H-burning ashes and it was engulfed in the upper layers of the convective He shell before the SN explosion. However, the bulk of the total $^{26}$Al yields is in the H-burning layers, in the mass region 4.4 M$_\odot$ < M < 4.8 M$_\odot$. Since the neutron-capture reactions on $^{26}$Al are not activated in H-burning conditions, for this specific stellar model we can expect that the impact of the $^{26}$Al(n, p)$^{26}$Mg and $^{26}$Al(n, α)$^{23}$Na uncertainties is marginal. The situation is completely different for the Z = 0.006 model: a sharp $^{26}$Al production peak is obtained at M = 3.35 M$_\odot$, along with the $^{28}$Si production due to explosive He-burning in the so-called C/Si zone (Pignatari et al. 2013). While the peak abundance of $^{26}$Al in these conditions is similar to the explosive C-burning peak, for the present model the C/Si zone is small in mass and therefore its contribution to the total $^{26}$Al yields is limited. The $^{26}$Al production due to H-burning is visible in the mass region 4.7 M$_\odot$ < M < 5.15 M$_\odot$, but with much smaller efficiency compared to the model at Z = 0.02. Therefore, the total $^{26}$Al yields are dominated by the (explosive) production in C-burning conditions, with a limited contribution from the H-burning component. In this case, we expect a significant impact of $^{26}$Al(n, p)$^{26}$Mg and $^{26}$Al(n, α)$^{23}$Na reactivities on $^{26}$Al yields.

From the earlier discussion, it becomes clear that the importance of $^{26}$Al(n, p)$^{26}$Mg and $^{26}$Al(n, α)$^{23}$Na reactions also depends on a number of properties developed during the evolution of stars. The differences discussed based on theoretical stellar simulations represents realistic variations also found in real stars. In models like in the upper panel of Fig. 8, the impact of the neutron capture rates on the $^{26}$Al yields would be small or negligible. On the other hand, for models like the M = 15 M$_\odot$ and Z = 0.006 (lower panel of the same figure) or the calculations shown by Iliadis et al. (2011) the uncertainties of the neutron capture rates on $^{26}$Al have a direct impact on the $^{26}$Al yields.

Notice that the two models considered here share the same initial progenitor mass and the same explosion energy setup (Ritter et al. 2018; Fryer et al. 2012), but they still show remarkable differences due to the intrinsic properties of the two stellar progenitors, developed during their evolution. While for the present models the main cause of such a different result is the change of initial metallicity, similar variations may be due to other initial parameters like the progenitor mass, rotation, or stellar binary interaction. More in general, we can also expect a non-linear impact of the uncertainty of these rates on the $^{26}$Al production with respect to the relevant stellar parameters mentioned earlier.

We now investigate the impact of the $^{26}$Al(n, p)$^{26}$Mg and $^{26}$Al(n, α)$^{23}$Na ground state rates and their uncertainties (Tables 1 and 2) on $^{26}$Al yields of the 15 M$_\odot$ stellar model with Z = 0.006, where the bulk of $^{26}$Al comes from explosive C-burning. Uncertainties in $^{26}$Al(n, p)$^{26}$Mg and $^{26}$Al(n, α)$^{23}$Na reactivities become progressively larger with increasing stellar temperature, due to experimental data being mainly available at lower neutron energies (see Tables 1 and 2, and Figs 2 and 4, respectively). Nucleosynthesis calculations were performed using the NuGrid post-processing network code PPN (e.g. Pignatari & Herwig 2012). The explosive single-zone trajectory was extracted from the M = 15 M$_\odot$ Z = 0.006 star shown in Fig. 8, at mass coordinate of M = 2.66 M$_\odot$, where the largest production of $^{26}$Al is obtained. The local temperature and density peaks during the CCSN explosion are 2.39 GK and 1.18 × 10$^5$ g cm$^{-3}$. In Fig. 9, the abundance evolution in the CCSN explosive trajectory is shown for two cases, using both the upper limits and the lower limits of the new $^{26}$Al neutron capture rates together. The other nuclear reactivities, including (n, p) and (n, α) reactivities on thermally excited states of $^{26}$Al, are not changed in the simulations (at peak temperatures of 2.39 GK, around 10 per cent of $^{26}$Al nuclei are in thermally excited states). The final $^{26}$Al abundance in mass fraction is varying by about a factor of 2.4, between 1.63 × 10$^{-4}$ and 6.87 × 10$^{-5}$. Notice how the $^{26}$Al abundance obtained by Ritter et al. (employing the $^{26}$Al(n, p)$^{26}$Mg and $^{26}$Al(n, α)$^{23}$Na from Caughlan & Fowler (1988) and the NACRE compilation (Angulo et al. 1999), respectively) is very close to what is obtained with our lower limits, meaning that the $^{26}$Al abundance decreases with our new rates. This is the opposite of what happens in AGB stars. Looking at Figs 2 and 4 it is possible to interpret this, as our new rates are higher than the older rates at high temperatures typical of CCSN explosions. We can consider the variation shown in Fig. 9 as a qualitative upper limit of the impact on the $^{26}$Al production on CCSNe due to the $^{26}$Al$(n, p)^{26}$Mg and $^{26}$Al$(n, \alpha)^{23}$Na uncertainties. Indeed, as we discussed earlier, in more extreme stellar conditions at higher temperatures $^{26}$Al is destroyed. For lower temperatures, our rate uncertainties (Tables 1 and 2) are smaller (in hydrostatic C-burning), or neutron reactions are not relevant (for H-burning conditions). To reduce uncertainties of $^{26}$Al yields from carbon burning environments in massive stars, new experimental data of $^{26}$Al$(n, p)^{26}$Mg and $^{26}$Al$(n, \alpha)^{23}$Na reaction cross-sections at higher neutron energies (hundreds of keV) are required.

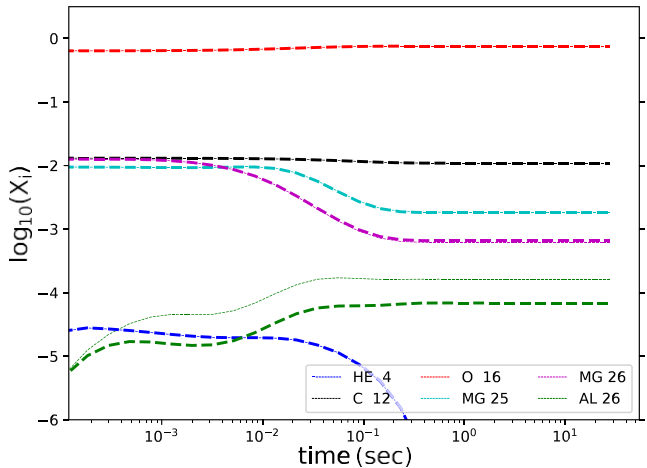

**Figure 9.** The evolution of the isotopic abundances (mass fractions) $^4$He, $^{12}$C, $^{16}$O, $^{25}$Mg, $^{26}$Mg, and $^{26}$Al are shown during the CCSN explosion using the $^{26}$Al$(n, p)^{26}$Mg and $^{26}$Al$(n, \alpha)^{23}$Na lower limit rates (thin lines) and upper limit rates (thick lines), respectively.

The overall abundance of $^{26}$Al observed via $\gamma$-rays in the Galaxy is built up by the total mass yield of $^{26}$Al ejected by CCSNe. This yield will be affected by the rate uncertainties investigated here by less than the factor of 2.4 reported earlier given that the total yields result from the sum of all the different mass regions. The composition of stardust SiC grains from CCSNe, instead, needs to be compared to the local abundances at each mass region of the ejecta because the grains are more likely to form from local rather than mixed ejecta material, see discussion in, for example, den Hartogh et al. (2022). These authors also confirmed that the standard CCSN models under-produce the $^{26}$Al/$^{27}$Al in stardust SiC grains from CCSNe. This picture is not changed substantially by using our new rates and their uncertainties and we confirm that different H mixing and burning processes appear to be required to match the data (Pignatari et al. 2013).

## 4 CONCLUSIONS

We presented new reactivities for the $^{26}$Al$(n, p)^{26}$Mg and $^{26}$Al$(n, \alpha)^{23}$Na nuclear reactions and tested their effect on stellar nucleosynthesis. We found that the new rates have a significant impact on both low-mass AGB and massive stars nucleosynthesis.

At temperatures relevant to AGB models (roughly up to 0.3 GK), the new rates are lower than those previously available and result in higher final $^{26}$Al/$^{27}$Al at the stellar surface. While stardust SiC grain data and model predictions are in broad agreement, a detailed comparison and robust conclusions are still hampered by systematic uncertainties present in the SiC data, in the determination of the temperature in the stellar models, and in the rate of the $^{22}$Ne($\alpha$, $n)^{25}$Mg reaction. Concerning CCSN nucleosynthesis in massive stars, we discussed the large stellar uncertainties still involved in the production of $^{26}$Al. Nevertheless, nuclear reactivities are also crucial to constrain the final $^{26}$Al yields. In particular, for CCSN models with a relevant explosive C/Ne burning component ejected, we showed that the $^{26}$Al abundance varies by up to a factor of 2.4 at the point in mass of highest production when adopting the upper or lower limit of our rates. This means that the total ejected yields will be affected by less than a factor of 2.4, since they result from the sum of all the different mass regions. Additionally, we confirm the conclusions from den Hartogh et al. (2022), who showed how standard CCSN models underproduce the $^{26}$Al/$^{27}$Al in stardust SiC grains from CCSNe. This result is still valid when our our new rates are adopted, and we confirm that different H mixing and burning processes appear to be required to match the data.

An improvement of the uncertainties from T9∼0.3 to 2.5 is required for future studies. A new measurement of these important reactions at high-neutron energy is planned with a new setup at the n_TOF CERN facility in the near future (Lederer-Woods et al. 2022).

## ACKNOWLEDGEMENTS

We thank Nan Liu, Peter Hoppe, and Fiorenzo Vincenzo for helpful discussions on the SiC grain data. For the purpose of open access, the author has applied a Creative Commons Attribution (CC BY) licence to any Author Accepted Manuscript version arising from this submission. UB and CLW acknowledge support from the European Research Council (grant agreements ERC-2015-STG number 677497). CLW and PJW acknowledge support from the Science and Technology Facilities Council UK (ST/M006085/1, ST/V001051/1). MP and ML thank the support from the ERC Consolidator Grant (Hungary) programme (RADIOSTAR, grant agreements number 724560). UB and MP acknowledge the support to NuGrid from JINA-CEE (NSF grant number PHY-1430152) and STFC (through the University of Hull's Consolidated Grant ST/R000840/1), and ongoing access to viper, the University of Hull High Performance Computing Facility. MP acknowledges the support from the 'Lendü¨let-2014" programme of the Hungarian Academy of Sciences (Hungary). DV acknowledges financial support from the German-Israeli Foundation (GIF number I-1500-303.7/2019). We thank the ChETEC COST Action (CA16117), supported by the European Cooperation in Science and Technology. This work was supported by the European Union's Horizon 2020 research and innovation programme (ChETEC-INFRA – Project number 101008324), and the IReNA network supported by US NSF AccelNet.

## DATA AVAILABILITY

The $^{26}$Al$(n, p)^{26}$Mg and $^{26}$Al$(n, \alpha)^{23}$Na reactivities presented in this work are available in machine-readable format table from the ChANUREPS nuclear platform (http://chanureps.chetec-infra.eu/chanureps/). The explosive single-zone trajectory, detailing the evolution of temperature and density over time, used to investigate the impact of our new rates on CCSN $^{26}$Al yields is available from the ORChESTRA platform (https://zenodo.org/communities/chetec-infra-wp4/). All the stellar models presented in this article will be shared on reasonable request to the corresponding author.

This paper has been typeset from a T$_E$X/L$^A$T$_E$X file prepared by the author.